\documentclass[a4paper]{jpconf}
\usepackage{graphicx}
\usepackage{iopams} 
\usepackage{braket}

\newcommand{\be}{\begin{equation}}
\newcommand{\ee}{\end{equation}}
\newcommand{\bea}{\begin{eqnarray}}
\newcommand{\eea}{\end{eqnarray}}
\newcommand{\non}{\nonumber}

\begin{document}
\title{Exact quantum dynamics of yrast states in the finite 1D Bose gas} 

\author{Eriko Kaminishi$^1$ , Jun Sato$^2$, Tetsuo Deguchi$^2$}
\address{
$^1$ Department of Physics, 
Graduate School of Science, 
The University of Tokyo, \\ 
7-3-1 Hongo, Bunkyo-ku, Tokyo 113-0033, Japan} 
\address{$^2$ Department of Physics, 
Graduate School of Humanities and Sciences, \\ 
Ochanomizu University, 2-1-1 
Ohtsuka, Bunkyo-ku, Tokyo 112-8610, Japan}

\ead{kaminishi@cat.phys.s.u-tokyo.ac.jp} 

\begin{abstract} 
We demonstrate that the quantum dynamics of yrast states in 
the one-dimensional (1D) Bose gas gives an 
illustrative example to equilibration 
of an isolated quantum many-body system. 
We first formulate the energy spectrum of yrast states in terms of 
the dressed energy by applying the method of finite-size corrections.     
We then review the exact time evolution of quantum states constructed from 
yrast states shown by the Bethe ansatz.   
In time evolution the density profile of an   
initially localized quantum state constructed from yrast states 
collapses into a flat profile in the case of a large particle number 
such as $N=1000$, while recurrence of the localized state occurs 
in the case of a small particle number such as $N=20$. 
We suggest that the dynamical relaxation behavior for the large $N$ case 
is consistent with the viewpoint of typicality for generic quantum states: 
the expectation values of local operators evaluated in most of quantum states 
are very close to those of the micro-canonical ensemble.     
%
\end{abstract}

\section{Introduction}

Ultracold quantum gases have attracted much interest due to their 
potential application to the testing of the quantum fluctuations in 
many-body systems \cite{Feshbach,Ketterle,Esslinger,BDZ08}. Furthermore, it has been shown that they are also quite useful for studying 
nontrivial dynamical properties 
of isolated quantum interacting systems 
not only theoretically \cite{Rigol} 
but also experimentally  \cite{Kinoshita,prethermal} (see for reviews,  
Ref. \cite{Yukalov,Polkovnikov}). 
The nonequilibrium behavior of isolated systems is closely related to 
fundamental aspects of quantum statistical mechanics 
such as quantum ergodic theorem and the concept of typicality among 
quantum states \cite{Tasaki,Lebowitz,von Neumann,Reimann,Sugita}. 
In this report we illustrate 
 how the dynamics of yrast states in the one-dimensional (1D) 
 Bose gas gives quite an illustrative example 
consistent with typicality of quantum states in isolated quantum  
many-body systems.   

Let us now introduce yrast states. 
In quantum many-body systems such as rotating nuclei, 
it is often important to study such eigenstates 
that have the lowest energy among all the eigenstates 
with the same given angular momentum \cite{BP99}. 
In fact, they may  provide an insight 
into ensembles under rotation.  
We call the lowest-energy state of a given angular momentum 
the {\it yrast state}. 
Yrast states are studied in several quantum systems such as 
the quantum Hall systems \cite{GDC10,Haldane}.  
In two dimensions,  yrast states 
are associated with dislocations or quantum vortices \cite{BR99,Viefers}. 
In the Gross-Pitaevskii mean-field picture,  
if the angular momentum is given by an integral multiple of the number of particles, $N$, the vortex core is located on the rotating axis 
(the centered single-vortex state), while 
if not, it is not (an off-centered vortex state) \cite{BR99}.

In order to study the 1D bosons of cold atoms 
we consider the Lieb-Liniger model \cite{Lieb-Liniger,Lieb63,Ol98}. 
It has various applications in the study of many-body phenomena 
with a wide range from the quasicondensate regime \cite{Ketterle,1Dex,NIST} 
to the strongly interacting Tonks-Girardeau regime \cite{Kinoshita,TG}. 
Assuming the periodic boundary conditions (PBC),  
the 1D system of length $L$ is equivalent to 
 that of the circle of circumference $L$. 
The periodic geometry has been experimentally realized 
as a circular waveguide or elliptic trap \cite{1Dex,NIST}.  
In the 1D system under PBC, the state with a given angular momentum is equivalent  
to that of the corresponding momentum \cite{KCU10}.  
Furthermore, the yrast states in the 1D Bose gas  
are associated with one-hole excitations \cite{yrast}. 
It has been pointed out that the dispersion relation of yrast states  
is very close to that of dark solitons \cite{IT80}. 
Dark solitons are experimentally realized in 
1D trap \cite{dark-soliton-exp}. 
We have a conjecture that quantum states of a dark soliton in the 1D Bose gas  
should be constructed from the yrast states \cite{dark-soliton}. 

Some fundamental aspects of quantum statistical mechanics 
particularly for isolated many-body systems 
have attracted much interest quite recently \cite{Yukalov,Polkovnikov}, 
and have been studied from the viewpoint of typicality of quantum states 
\cite{Tasaki,Lebowitz,von Neumann,Reimann,Sugita}. 
Although one of the earliest exact studies of nonequilibrium dynamics of quantum many-body systems goes back to 1970s \cite{McCoy1970}, 
the problem of quantum quench has been investigated extensively 
through theoretical approaches quite recently during the last decade, 
partially motivated by the experimental realization of quantum many-body systems in cold atomic systems \cite{Calabrese-Cardy,SeiSuzuki,Mossel2010,EsslerPRL,Rigol2011}. 
Relaxation and thermalization phenomena 
have been observed not only theoretically in the study of quantum quench 
but also experimentally in cold atomic systems \cite{prethermal}.  
Thus, it is an interesting problem to study many-body phenomena in the 1D Bose gas theoretically and to compare the theoretical results with experimental ones.

It is one of the greatest achievements in exactly solvable models  
that finite-temperature specific heat and magnetic susceptibility 
are explicitly evaluated by the Bethe-ansatz method 
\cite{Korepin,M-Takahashi,1DHubbard}.  On the other hand, until quite recently 
it was not easy to evaluate correlation functions of quantum integrable 
systems even numerically for large systems. 
However, by combining  Slavnov's formula of scalar products 
\cite{Slavnov1989,Slavnov1990} 
with the formulas of the Quantum Inverse Scattering Problem \cite{Lyon}, 
or by improving some formulas of form factors for the 1D Bose gas 
\cite{Kojima,Caux-Calabrese-Slavnov2007,Calabrese-Caux2007}, 
one can  evaluate numerically correlation functions and form factors of 
some operators of the XXZ spin chain \cite{Karbach,JSato,Caux-Hagemans-Maillet} 
and the 1D Bose gas \cite{Caux-Calabrese-Slavnov2007,Calabrese-Caux2007}
by the Bethe ansatz method (for numerical values of finite-temperature XXX correlation functions, see Ref. \cite{Wuppertal}).

The main purpose of the present report is to review  
interesting dynamical properties in the quantum dynamics 
associated with yrast states in the 1D Bose gas \cite{SKKD1}. 
The contents of the present report consist of the following: 
In section 2 we express the energy spectrum of yrast states 
in terms of the dressed energy. 
We assume that the system size is finite for yrast states,   
and generalize the method of finite-size corrections 
in the the ground-state energy of the 1D Bose gas. 
Here some details given in  section 2 have not been presented in Ref. \cite{yrast}.     
In section 3 we review that some linear combination of yrast states 
leads to a quantum state which is localized in terms of the density operator, i.e., 
it has a density notch in the density profile, and also that  
the density profile collapses into a flat profile in time evolution 
\cite{SKKD1}.

%
%
\section{Yrast energy spectrum via finite-size corrections in the 1D Bose gas}

\subsection{Lieb-Liniger model and the Bethe-ansatz method}

The Hamiltonian of the Lieb-Liniger (LL) model \cite{Lieb-Liniger} is given by 
\begin{equation}
{\cal H}_{\rm LL} = - {\frac {\hbar^2} {2m}} 
\sum_{j=1}^N {\frac {\partial^2} {{\partial x_j}^2}} 
+ g_{\rm 1D} \sum_{j < k} ^N\delta(x_j - x_k)  \, .  \label{eq:LL}
\end{equation}
Here $g_{\rm 1D}$ gives the 1D coupling constant. 
We assume the periodic boundary conditions: 
We identify the position of coordinate $x$ with that of $x+L$.

An exact eigenvector of the Hamiltonian~(\ref{eq:LL}) 
is derived by the Bethe-ansatz method, if a set of quasi-momenta 
$k_j$'s for $j \in \{1, 2, \ldots, N\}$  
satisfies the Bethe-ansatz equations (BAE): 
\be 
L k_j = 2\pi I_j -  \sum_{\alpha \ne j}^N 2 \tan^{-1} 
\left(\frac{k_j- k_{\alpha}}{c}\right)
\quad  \mbox{for } \quad j=1, 2, \ldots, N.  \label{eq:BAE}  
\ee
Here $c= m g_{\rm 1D}/\hbar^2$ is the coupling constant 
with the dimension of wavenumber, and quantum numbers $I_j$ 
are integers for odd $N$ and half-integers for even $N$. 

For the repulsive interaction: $c> 0$, it has been shown that 
the Bethe-ansatz eigenvectors are complete \cite{Dorlas} 
(for some generalization, see Ref. \cite{Emisz}). Therefore,   
each eigenvector of the Hamiltonian~(\ref{eq:LL}) 
is specified by a set of quantum numbers $I_j$. 
The ground-state solution, for instance, corresponds 
to the set of integers:  
$I_j^{\rm (g)} = j - (N+1)/2$ for $j \in \{1, 2, \ldots, N\}$. 
Hereafter we consider only the case of repulsive interaction.

Thus, by solving the BAE for a given set of $I_j$, 
we obtain a set of quasi-momenta $k_j$. 
The eigenvalue of the Hamiltonian~(\ref{eq:LL}) 
is given by the sum of $k_j^2$ over all $j$ \cite{Lieb-Liniger,Lieb63}. 
The total angular momentum is given by $Y= R \hbar \sum_{j=1}^{N} k_j$, 
or, according to the BAE, alternatively by   
$Y = \hbar \sum_{j=1}^{N} I_j$.

\subsection{Particle-hole excitations}

We derive an excitation with one hole and one particle 
from the ground state $I_j^{\rm (g)}$.  
We make a hole at $I_h$ and a particle at $I_p$. 
We denote by ${k}_j^{(1)}$ 
the quasi-momenta of the particle-hole excitation 
satisfying the Bethe ansatz equations (\ref{eq:BAE}) 
with integers $I_j^{(1)}$ for $j=1, 2, \ldots N-1$ and 
$I_p^{(1)}=I_p$, which are given by  
\be 
I_j^{(1)} =  I_j^{(g)} + H(j-j_h) \quad  {\rm for}  
\quad j=1, 2, \ldots, N-1 , 
\ee
and $I_p^{(1)} = I_p >  I_{N}^{(g)}$. 
Here integer $j_h$ denotes the hole position: $j_h= I_h + (N+1)/2$  
and the symbol $H(x)$ denotes the Heaviside step function:     
$H(x) = 1$ for $x \ge 0$, and $H(x) = 0$ otherwise. 

The angular momentum $Y^{p,h}$ of  
the particle-hole excited state with a particle at $I_p$ and 
a hole at $I_h$ is given by 
\be 
Y^{p,h}= \hbar (I_p  - I_h ) \, .    
\ee
If $Y^{p,h}$ is nonzero, 
the dispersion relation is not symmetric. 
It seems that it is not straightforward to apply the 
method of finite-size corrections to the spectrum of yrast states.  

Let us define the counting function 
$Z^{(1)}(k)$ for the particle-hole excitation by 
\be 
Z^{(1)}(k)  =  {\frac k {2 \pi}} + {\frac 1 L} \sum_{\alpha=1}^{N-1} 
\frac 1 {2 \pi} 2 \tan^{-1}\left({\frac {k-k_{\alpha}^{(1)}} c} \right) + 
 {\frac 1 L}  {\frac 1 {2 \pi}} 2 \tan^{-1} 
\left({\frac {k-k_{p}^{(1)}} c} \right) \, . 
\ee 
In terms of $Z^{(1)}(k)$ 
the Bethe-ansatz equations are written as follows: 
$Z^{(1)}(k_j^{(1)}) = I_j^{(1)}/L$ for 
$j=1, 2, \ldots, N-1$; $Z^{(1)}(k_p^{(1)}) = I_p^{(1)}/L$.   
We define a continuous counting function ${\hat Z}(k)$ by 
\be 
{\hat Z}(k) = Z^{(1)}(k) - {\frac 1 L} H(k-{\tilde k}_h) . 
\ee
Here we determine ${\tilde k}_h$ by ${\hat Z}({\tilde k}_h)=I_h/L$ 
\cite{yrast}. 
We introduce the Fermi parameters $Q^{\pm}$ by 
\be 
{\hat Z}(Q^{+}) = {\frac {I_{N-1}^{(1)} +1/2} L } \quad , \quad 
{\hat Z}(Q^{-}) = {\frac {I_{1}^{(1)} -1/2} L } \, .  \label{eq:FermiQ}
\ee 
We define the root density ${\hat \rho}(k)$ 
by the derivative of ${\hat Z}(k)$ with respect to $k$: 
${\hat \rho}(k) = {d {\hat Z}}/ {dk}$. 
Shifting ${\hat \rho(k)}$ as 
$\rho_{p,h}(k)= {\hat \rho(k)} + \delta(k-k_p^{(1)})/L$, 
through formula (\ref{eq:Euler-Maclaurin2}) we have 
\be 
\rho_{p,h}(k | Q^{\pm}) = \frac 1 {2 \pi} + 
{\frac 1 L} \left( \delta(k-k_p^{(1)}) - \delta(k-{\tilde k}_h) \right) 
+ \frac 1 {2 \pi L} K(k-k_p^{(1)}) 
+ \int_{Q^{-}}^{Q^{+}} 
 K(k-q) \rho_{p,h}(q | Q^{\pm}) {\frac {dq} {2\pi}} 
\label{eq:IEph}
\ee
Here, the integral kernel $K(k)$ 
is given by $K(k)=2c/(k^2+c^2)$.

%
%
\subsection{Dressed energy and formal solutions to the integral equations}

We define the dressed energy \cite{Korepin} by 
\begin{equation} 
\label{eq:epsilon}
\epsilon(k) = \epsilon_0(k) + 
\int_{Q^{-}}^{Q^{+}} {\frac 1 {2\pi}} K(k-q) \epsilon(q) dq  \, . 
\label{eq:dressed-energy}
\end{equation}
Here $\epsilon_0(k)$ denotes the bare energy. 
Let us define the integral operator ${\hat K}$ by 
\be 
\left( {\hat K} f \right)(k) = \int_{Q^{-}}^{Q^{+}} K(k-q) f(q) dq  \, . 
\ee
We express it also by $\langle k | {\hat K} f \rangle$. 
Writing the dressed energy and the bare energy 
as $\epsilon(k) =\langle k | \epsilon \rangle$ and 
$\epsilon_0(k)= \langle k | \epsilon_0 \rangle$, respectively,   
we formulate the integral equation (\ref{eq:dressed-energy}) as follows: 
$| \epsilon \rangle  = | \epsilon_0 \rangle + 
{\hat K} | \epsilon \rangle/2 \pi $.  
It has the solution of an infinite series:  
\be 
|\epsilon \rangle = |\epsilon_0 \rangle + 
\left( \frac 1 {2\pi} {\hat K} \right) |\epsilon_0 \rangle 
+ \left( \frac 1 {2\pi} {\hat K} \right)^2 |\epsilon_0 \rangle + \cdots 
=  {\frac I {I -  \frac 1 {2\pi} {\hat K}}} |\epsilon_0 \rangle \, . 
\ee
Here $I$ denotes the identity operator. We define the bulk part of the root density of 
the particle-hole excitation, $\rho(k|Q^{\pm})$, by the solution of the integral equation: 
\be 
\rho(k| Q^{\pm}) = {\frac 1 {2 \pi}} + \int_{Q^{-}}^{Q^{+}} 
{\frac 1 {2 \pi}} K(k-q)  \rho(q|Q^{\pm}) dq \, . 
\ee
Let us express the constant function of variable $k$ 
as  $\langle k | I \rangle = 1 $. 
Then, the root density $\rho(k|Q^{\pm})$ is written formally as  
\be 
\rho(k| Q^{\pm}) = 
{\frac 1 {2 \pi}} \langle k |  
{\frac I {I -  \frac 1 {2\pi} {\hat K}}} |I \rangle  \, . 
\ee
Thus, the bulk part of the particle-hole excitation energy  
is expressed in terms of the dressed energy as 
\be 
\int_{Q^{-}}^{Q^{+}} \epsilon_0(k) \rho(k|Q^{\pm}) {dk}
 = \int_{Q^{-}}^{Q^{+}} \epsilon(k)  {\frac {dk} {2 \pi}} \, . 
\label{eq:DE}
\ee

Let us now solve integral equation (\ref{eq:IEph}).  
We define function $\eta(k; q|Q^{\pm})$ 
by the solution of the integral equation: 
\be 
\eta(k; q|Q^{\pm}) = \delta(k-q) + \int_{Q^{-}}^{Q^{+}} 
{\frac 1 {2 \pi}} K(k-k^{'})  \eta(k^{'}; q |Q^{\pm}) dk^{'} \, . 
\ee
Formulating the solution of eq. (\ref{eq:IEph}) 
as the sum of the bulk density $\rho(k|Q^{\pm})$ and corrections  
$\Delta \rho_{p,h}(k| Q^{\pm})$:  
$\rho_{p,h}(k| Q^{\pm})  = \rho(k|Q^{\pm}) 
+ \Delta \rho_{p,h}(k | Q^{\pm})$,  we have 
\be 
\Delta \rho_{p,h}(k | Q^{\pm})  = 
{\frac 1 L} \left( \delta(k-k_p^{(1)}) 
+ \int_{Q^{-}}^{Q^{+}} \eta(k; k^{'}|Q^{\pm}) 
{\frac 1 {2\pi}}K(k^{'}-k_p^{(1)}) dk^{'} 
- \eta(k; {\tilde k}_h| Q^{\pm}) \right)  \, .   
\label{eq:Delta-rho}
\ee

%
%
\subsection{Energy spectrum of the particle-hole excitation }

We introduce two Lagrange multipliers, 
chemical potential $\mu$ and angular velocity $\Omega$, respectively.   
We shift the LL Hamiltonian as 
${\cal H}_{Y, \, N} = {\cal H}_{\rm LL} - \mu N - \Omega Y$.   
We now define the dressed energy $\epsilon(k)$ by eq. (\ref{eq:dressed-energy}) 
with the new bare energy $\epsilon_0 (k)= (\hbar k)^2 /(2m) - \mu - \Omega R\hbar k $, 
where the Fermi parameters $Q^{\pm}$ are given by eqs. (\ref{eq:FermiQ}).   
We determine parameters $\mu$ and $\Omega$ by the following conditions: 
$\epsilon(Q^{+}) = \epsilon(Q^{-})=0$.

The energy of the particle-hole excitation with a particle at $I_p$ 
and a hole at $I_h$ is given by  
\be 
E_{Y, N}^{p,h} =  
\sum_{\alpha=1}^{N-1} \epsilon_0 (k_{\alpha}^{(1)}) + 
\epsilon_0(k_p^{(1)}) \, . 
\ee
For large system size $L$, the particle-hole excitation energy
$E_{Y, N}^{p,h}$ is expressed in terms of the root density 
through formula (\ref{eq:Euler-Maclaurin2}).  
\bea 
E_{Y, N}^{p,h} 
& = & L \int_{Q^{-}}^{Q^{+}} \epsilon_0(k) \rho_{p,h}(k|Q^{\pm}) dk 
+ \epsilon_0(k_p^{(1)}) \non \\ 
& = & 
L \int_{Q^{-}}^{Q^{+}} \epsilon_0(k)  \rho(k|Q^{\pm}) dk 
+ \left( \epsilon_0(k_p^{(1)}) + 
L \int_{Q^{-}}^{Q^{+}} \epsilon_0(k) \Delta \rho_{p,h}(k|Q^{\pm}) dk \right) . 
\eea
Here we recall $\rho_{p,h}(k|Q^{\pm}) 
= \rho(k|Q^{\pm}) + \Delta \rho_{p,h}(k|Q^{\pm})$.   
With the solution (\ref{eq:Delta-rho}) we have  
\be 
\epsilon_0(k_p^{(1)}) 
+ L \int_{Q^{-}}^{Q^{+}} 
\epsilon_0 (k) \Delta \rho_{p,h}(k | Q^{\pm}) dk = 
\epsilon(k_p^{(1)}) - \epsilon({\tilde k}_h) \, .  
\ee
Thus, through relation (\ref{eq:DE})  
we express the particle-hole excitation energy $E_{Y, N}^{p, h}$ 
in terms of the dressed energy 
 as follows. 
\begin{equation}
E_{Y, N}^{p, h } = L \int^{Q^{+}}_{Q^{-}} \epsilon(k) 
{\frac {dk} {2 \pi}} 
+ \epsilon({k}_p^{(1)})  - \epsilon({\tilde k}_h) 
- {\frac {\pi v_F} {6 L}} +O(L^{-2}) . \label{eq:EYN}
\end{equation}

The exact expression (\ref{eq:EYN}) of the asymptotic expansion 
with respect to the inverse of system size $L$ contains both 
the Lieb's type I and type II excitations \cite{Lieb63} together with 
the finite-size corrections to the yrast energy  
of a given angular momentum $Y^{p,h}$ \cite{yrast}.  
The spectrum of yrast excited states are shown 
by making use of expression (\ref{eq:EYN}) in Ref. \cite{yrast}.   

%
%
\section{Non-equilibrium dynamics of localized quantum states}

\subsection{Second quantized Hamiltonian of the Lieb-Liniger model}

Let us employ the unit system with $2m=\hbar =1$. 
%
In terms of the canonical Bose field ${\hat \psi}(x,t)$  
the second quantized Hamiltonian of 
the Lieb-Liniger model is given by 
\be
{\cal H}_{\rm NLS} = 
\int_{0}^{L} dx [ \partial_x {\hat \psi}^{\dagger} \partial_x {\hat \psi} + 
c {\hat \psi}^{\dagger} {\hat \psi}^{\dagger}  {\hat \psi} {\hat \psi} 
] . 
\ee
The second quantized Hamiltonian ${\cal H}_{\rm NLS}$ leads to 
the quantum nonlinear Schr{\"o}dinger equation:  
$i \partial_t {\hat \psi} =  - \partial^2_x {\hat \psi}  
+ 2c {\hat \psi}^{\dagger} {\hat \psi} {\hat \psi}$. 


The bulk properties of the LL model are characterized by a single parameter $\gamma=c/n$, where $n=N/L$ is the number density of particles. 
We fix the particle density as $n=1$ and vary the coupling constant $c$. 
The unit time in our simulation is proportional to $L^{-2}$.

%
%
\subsection{Exact time evolution of density notch: collapse into a flat profile} 
 
Let us construct an initially localized quantum state \cite{SKKD1}. 
We denote by $|P \rangle$ the one-hole excitation 
with total momentum $P=2 \pi p/L$. Here, integer $p$ takes 
an integer value from the set $\{0, 1, \ldots, N-1\}$. 
It corresponds to the particle-hole 
excitation with $I_h= (N-1)/2 - p$ and $I_p= (N+1)/2$.  
Here, integer $q$ satisfies $0 \le q \le N-1$. 
We define quantum states with a density notch, $|X \rangle$, 
by the discrete Fourier transformation of $|P \rangle$ as follows.  
\begin{equation}
| X \rangle = \frac 1 {\sqrt{N}} \sum_{p=0}^{N-1} 
\exp(- 2 \pi i p q/N) | P \rangle \, . 
\label{xket}
\end{equation}
Here, $|X \rangle$ has the coordinate $X=qL/N$, and
its density profile has a density notch at position $x$ with $x=X+L/2$. 

\begin{figure}
\begin{center}
\includegraphics[width=0.35\columnwidth]{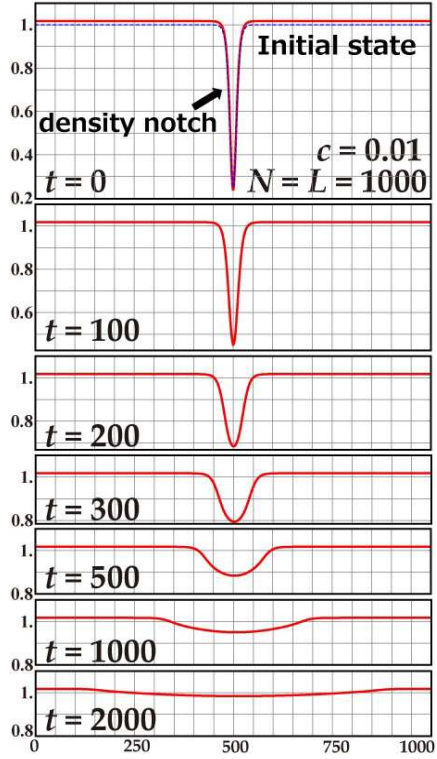}
\end{center}
\caption{
Snapshots of the exact time evolution of density profile $\rho(x,t)$ 
with $c=0.01$ for $N=1000$, $L=1000$ (red solid line).
}
\label{snap_shots_c001}
\end{figure}

In order to study the dynamics of the 1D Bose gas, we evaluate form factors 
of the density operator of the 1D Bose gas.
Hereafter we consider the Schr{\" o}dinger picture. 
The time evolution of the quantum state is given by 
$|X(t)\rangle = \exp(-i {\cal H}t)|X\rangle$. 
We evaluate the expectation value of the density operator 
$\hat{\rho}(x)=\hat{\psi}^\dagger(x) \hat{ \psi}(x)$ 
with respect to state $|X(t) \rangle$ at time $t$  as 
\begin{equation}
\langle X(t)|\rho(x)| X(t)\rangle
=\sum^{N-1}_{p,p'=0}e^{-2\pi i(p-p')q/N} 
e^{i(P-P')x-i(E_p-E_{p'})t}
\langle P'|\rho(0)|P\rangle, \label{eq:rho(t)}
\end{equation}
Here we recall that $P=2\pi p/L$ and $P'=2\pi p'/L$ denote the total momenta.  
The energy eigenvalues $E_p$ of type-II excitations are readily 
evaluated by solving the BAE numerically.

We calculate the form factors $\langle P'|\rho(0)|P\rangle$ 
in eq. (\ref{eq:rho(t)}) by making use of Slavnov's formula 
\cite{Slavnov1989,Slavnov1990} and the Gaudin-Korepin norm formula 
\cite{Korepin}. Slavnov's formula is given by 
\begin{equation}
\langle{P'}|\rho(0){|P}{\rangle}=i^{N}(P-P')
(\prod^N_{j,\ell} \frac{k_{j}-k_{\ell}+ic}{k'_{j}-k_\ell}) \, 
{\rm det}_{N-1} \, U(k,k'), \label{eq:Slavnov}
\end{equation}
where pseudomomenta $\{k_1,\cdots,k_N\}$ and 
$\{k'_1,\cdots,k'_N\}$ correspond to 
$|P\rangle$ and $|P'\rangle$, respectively, and 
the matrix elements of an $(N-1) \times (N-1)$ matrix $U(k,k')$ 
are given for $j, l= 1, 2, \ldots, N-1$, as follows. 
\begin{equation}
U(k,k')_{j,\ell} =2\delta_{j\ell} {\rm Im} 
\left[ \prod^N_{a=1}
\frac{k'_a-k_j + ic}{k_a-k_j + ic} \right]
+\frac{\prod^N_{a=1}(k'_a-k_j)}{\prod^N_{a\neq j}(k_a-k_j)}
\left( K(k_j-k_\ell)-K(k_N-k_\ell) \right) \, . 
\label{eq:matrixU}
\end{equation}

It is shown in Fig. 1 that the density notch in the initial state collapses into 
a flat profile, which gives the equilibrium state in the density profile. 
The collapse of the density notch shows 
relaxation, or more precisely, equilibration of the expectation value of 
the density operator, which occurs in most of 
isolated quantum many-body systems due to typicality of quantum states.   
Here, from the typicality viewpoint, we assume that  
most of quantum states are close to be in equilibrium: 
the expectation values of local operators 
in most of quantum states are very close 
to those of the micro-canonical ensemble.  

More intuitively, we can explain it as follows. 
Suppose that a given quantum state corresponds to a representative point in the phase space and it moves in the phase space during time evolution. 
Then, the trajectory starting from a nonequilibrium quantum state passes through many equilibrium quantum states, since the majority of quantum states are close to be in equilibrium. It thus follows that the expectation value of a local physical quantity approaches that of the (micro-)canonical ensemble.  

We now  give some comments on relevant researches. 
In the weak coupling region the correlation length (healing length) of the 1D Bose gas 
increases as the coupling constant $\gamma$ approaches zero. 
Here, quite a novel finite-size scaling behavior of condensate fraction holds \cite{FS-BEC}.  
For a large system with $L$ being infinite, the collapse of an initially localized density profile can also be considered as the collapse of an initial wave packet due to the nonlinear dispersion relation of matter waves.  Here we remark that coherent states can be  constructed in a finite system size $L$ \cite{Anton}. 
Typicality of quantum states should be associated with 
the eigenstate thermalization hypothesis, which is studied also in the 1D Bose gas 
 \cite{Ikeda}.   Quite recently, relaxation time is rigorously evaluated for 
typical isolated quantum systems \cite{Hara}.

\subsection{Recurrence of the density notch}

For the small number of particles such as $N=20$, 
we observed recurrence phenomenon \cite{SKKD1}. 
In particular, in the free-fermionic and the free-bosonic regimes, 
there are many quantum states which show recurrence \cite{Recurrence}.

Finally,  we remark that the quantum system we have investigated has 
a finite number of particles, not an infinite one. 
Thus, the relaxation behavior observed in Fig. 1 is not necessarily complete. 
It might recur after a very long period of time such as the age of the universe.

\begin{figure}
\begin{center}
\includegraphics[width=0.31\columnwidth]{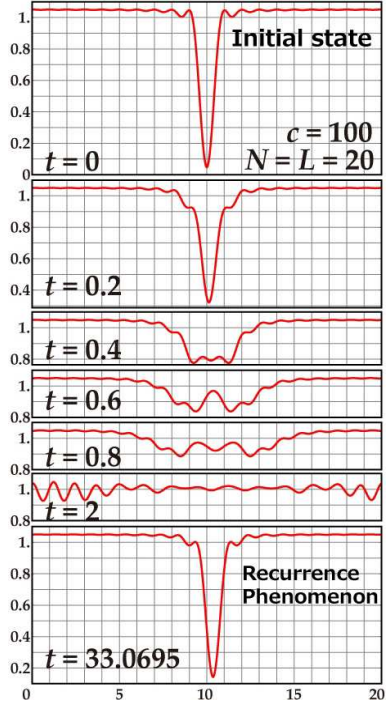}
\end{center}
\caption{
Snapshots of density profiles in time evolution after  
the initial state constructed from yrast states 
($c=100$ , $N=20$, $L=20$). The density notch appears again 
at $t=33.0695$. 
}
\label{snap_shots_N20}
\end{figure}

\section*{Acknowledgment}
The authors would like to thank H. Fujishima, R. Hatakeyama, R. Kanamoto 
and M. Ueda for useful comments. 
This work was partially supported by 
by Grant-in-Aid for Scientific Research No. 24540396.  
E. K. is supported by JSPS.

\appendix 
\section{The Euler-Maclaurin formula}
Making use of the Euler-Maclaurin formula for any given analytic function $f(x)$ 
\be 
{\frac 1 L} \sum_{j=n_1}^{n_2} f(j/L) = 
\int_{(n_1- 1/2)/L}^{(n_2+1/2)/L} f(x) dx 
- {\frac 1 {24 L^2}} \left( {\frac {df} {dx}}({\frac {n_2+1/2} L}) - 
{\frac {df} {dx}} ({\frac {n_1- 1/2} L}) \right) + o({\frac 1 {L^2}}) ,  
\label{eq:Euler-Maclaurin}
\ee   
we can approximate the sum of $f(k_j)$ over an interval  
of integer $j$ as an integral of $k$:   
\be 
{\frac 1 L} \sum_{\alpha=1}^{N-1} f(k_{\alpha}^{(1)}) = 
\int_{Q^{-}}^{Q^{+}} f(k) {\hat \rho}(k) dk \, + O(1/L). 
\label{eq:Euler-Maclaurin2}
\ee

\end{document}